\documentclass[nofootinbib,preprint,tightenlines]{revtex4}

\usepackage{graphicx}
\usepackage{amsmath}
\usepackage{amssymb}

\newcommand{\beq}{\begin{eqnarray}}
\newcommand{\eeq}{\end{eqnarray}}
\newcommand{\bqa}{\begin{eqnarray}}
\newcommand{\eqa}{\end{eqnarray}}

\newcommand{\sing}{^1{\rm S}_0}
\newcommand{\trip}{^3{\rm S}_1}

\newcommand{\ii}{\mathrm{i}}
\newcommand{\e}{\mathrm{e}}
\newcommand{\dd}{\mathrm{d}}
\newcommand{\intdd}[1]{\mathrm{d}^3{#1}\,}
\newcommand{\intdpo}[1]{\frac{\mathrm{d}{#1}}{(2\pi)}\,}

\newcommand{\halb}{\frac{1}{2}}
\newcommand{\lp}{{\ell^{\prime}}}
\newcommand{\lpp}{{\ell^{\prime\prime}}}
\renewcommand{\mp}{{m^{\prime}}}
\newcommand{\mpp}{{m^{\prime\prime}}}

\def\mqo2{{\!\!\!}}

\begin{document}


\preprint{HISKP-TH-09/32}
\title{On the modification of the Efimov spectrum in a finite cubic box}
\author{Simon Kreuzer}

\author{H.-W. Hammer}
\affiliation{Helmholtz-Institut f\"ur Strahlen- und Kernphysik (Theorie)\\
and Bethe Center for Theoretical Physics,
 Universit\"at Bonn, 53115 Bonn, Germany\\}

\date{October 12, 2009}

\begin{abstract}
  Three particles with large scattering length display a universal 
  spectrum of three-body bound states called \lq\lq Efimov trimers''. 
  We calculate the modification of the Efimov trimers of three identical bosons
  in a finite cubic box and compute the dependence of
  their energies on the box size using effective field theory.
  Previous calculations for positive scattering length that were 
  perturbative in the finite volume energy shift are extended to 
  arbitrarily large shifts and negative scattering lengths.
  The renormalization of the effective field theory in the finite volume
  is explicitly verified. We investigate the effects of partial
  wave mixing and study the behavior of shallow trimers near the dimer
  energy. Moreover, we provide numerical evidence for universal scaling
  of the finite volume corrections.
\end{abstract}

\maketitle

\section{Introduction}

Few-body systems with resonant interactions characterized by a large
scattering length $a$ show interesting universal properties.
If $a$ is positive, 
two particles of mass $m$ form a shallow dimer with energy
$E_2 \approx -{\hbar^2}/{(m a^2)}\,$,
independent of the mechanism responsible for the large scattering length.
Examples for such shallow dimer states are the deuteron 
in nuclear physics, the $^4$He dimer in atomic physics, and 
possibly the new charmonium state $X(3872)$ in particle 
physics \cite{Braaten:2004rn,Platter:2009gz}.  
In the three-body system, the universal properties include the
Efimov effect \cite{Efimov-70}.  If at least two of the three pairs of
particles have a large scattering length $|a|$ compared to the range
$r_0$ of their interaction, there is a sequence of three-body bound
states whose energies are spaced geometrically between
$-\hbar^2 / m r_0^2$ and $-\hbar^2 / m a^2$.  In the limit $1/a \to 0$, there
are infinitely many bound states with an accumulation point at
the three-body scattering threshold. These Efimov states or trimers have a
geometric spectrum \cite{Efimov-70}:
\begin{eqnarray}
E^{(n)}_3 = -(e^{-2\pi/s_0})^{n-n_*} \hbar^2 \kappa^2_* /m,
\label{kappa-star}
\end{eqnarray}
which is specified by the binding momentum $\kappa_*$ of the Efimov trimer
labeled by $n_*$. This spectrum is a consequence of a discrete scaling
symmetry with discrete scaling factor $e^{\pi/s_0}$.  In the case of
identical bosons, $s_0 \approx 1.00624$ and the discrete scaling
factor is $e^{\pi/s_0} \approx 22.7$.  The discrete
scale invariance persists if $a$ is large but finite, but in this 
case it connects states corresponding to different values of
the scattering length.  The scaling symmetry becomes 
also manifest in the log-periodic dependence of 
scattering observables on the scattering length $a$ \cite{Efimov79}.  The
consequences of discrete scale invariance and \lq\lq Efimov physics'' can be
calculated in an effective field theory for short-range interactions,
where the Efimov effect appears as a consequence of a renormalization
group limit cycle \cite{Bedaque:1998kg}.

While the Efimov effect was established theoretically already in 1970,
first experimental evidence for an Efimov trimer in ultracold Cs atoms was
provided only recently by its signature in the three-body recombination rate
\cite{Kraemer-06}. It could be unravelled by varying the
scattering length $a$ over several orders of magnitude using a
Feshbach resonance.  Since this pioneering experiment, there was 
significant experimental progress in observing Efimov physics 
in ultracold quantum gases.
More recently, evidence for Efimov trimers was
also obtained in atom-dimer scattering \cite{Knoop08} and in
three-body recombination in a balanced mixture of atoms in three
different hyperfine states of $^6$Li \cite{Ottenstein08,Huckans08},
in a mixture of Potassium and Rubidium atoms \cite{Barontini09},
and in an ultracold gas of $^7$Li atoms \cite{Gross09}. In another experiment
with Potassium atoms \cite{Zaccanti09}, two bound trimers were observed.

The observation of Efimov physics in nuclear and particle physics
systems is complicated by the inability to vary the scattering length
and one has to focus on the detection of excited states.
While two-neutron halo nuclei could be bound due to the Efimov effect,
the analysis of known halo nuclei does not show much promise for an 
unambiguous identification (See Ref.~\cite{Canham:2008jd} and references
therein). Another opportunity to observe Efimov physics
is given by lattice QCD simulations of three-nucleon systems
\cite{Wilson:2004de}. A number of studies of the quark-mass
dependence of the chiral nucleon-nucleon ($NN$) interaction found that
the inverse scattering lengths in the relevant $\trip$--$^3{\rm D}_1$
and $\sing$ channels may both vanish if one extrapolates away from the
physical values to slightly larger quark masses
\cite{Beane:2001bc,Beane:2002xf,Epelbaum:2002gb}. This implies that
QCD is close to the critical trajectory for an infrared renormalization group
limit cycle in the three-nucleon sector.  It was conjectured that QCD could 
be tuned to lie precisely on the critical trajectory by tuning the up and
down quark masses separately \cite{Braaten:2003eu}.
As a consequence, the triton would display the Efimov effect. More
refined studies of the signature of Efimov physics in this case
followed \cite{Epelbaum:2006jc,Hammer:2007kq}. However, a proof of
this conjecture can only be given by an observation of this effect in a
lattice QCD simulation \cite{Wilson:2004de}. The first full lattice
QCD calculation of nucleon-nucleon scattering was reported in
\cite{Beane:2006mx} but statistical noise presented a serious challenge.
A promising recent high-statistics study of three-baryon systems 
presented also initial results for a system with the quantum numbers 
of the triton such that lattice QCD calculations of three-nucleon 
systems are now within sight
\cite{Beane:2009gs}. For a review of these activities, see
Ref.~\cite{Beane:2008dv}. Since
lattice simulations are carried out in a cubic box, it is important to
understand the properties of Efimov states in the box. The first step
towards this goal is to understand these modifications for a system
of three identical bosons.

The corresponding modifications of the Efimov spectrum can be calculated in
effective field theory (EFT) since the finite volume modifies the
infrared properties of the system. 
The properties of two-body systems with large scattering length in
a cubic box were calculated in Ref.~\cite{Beane:2003da}.
Some properties of three-body
systems in a finite volume have also been studied previously. For
repulsive and weakly attractive interactions without bound states, Tan
has determined the volume dependence of the ground state energy of
three bosons up to~$\mathcal{O}((a/L)^7)$ \cite{Tan08}.  In
Refs.~\cite{Beane:2007qr,Detmold:2008gh}, this result was extended for
general systems of $N$ bosons. It was used to
analyze recent results for three and more boson systems from lattice QCD 
\cite{Beane:2007es,Detmold:2008fn,Detmold:2008yn}.
For the unitary limit of infinite scattering length, some studies
have been carried out as well.
The properties of three spin-1/2 fermions in a box were 
investigated in \cite{Pricoup07}. However, this system has no
three-body bound states in the infinite volume.
The volume dependence of energy levels in a finite volume can also
be used to extract scattering phase shifts and resonance properties
from lattice calculations \cite{Luscher:1990ux,Luscher:1991cf}.
For a recent application of this idea to the $\Delta(1232)$ resonance,
see Refs.~\cite{UGM1,UGM2}.

In a previous letter, we have investigated the finite volume corrections for a
three-boson system with large but finite scattering length
\cite{Kreuzer:2008bi}. We have studied the modification
of the bound state spectrum for positive scattering length
using an expansion valid for small finite volume shifts
and explicitly verified the renormalization in the finite box
within this approximation. In the current paper, 
we present an extension of this work that avoids the expansion in the 
energy shift and can be applied to arbitrarily large shifts. Moreover,
we extend our previous studies to the case of negative scattering lengths. Our results indicate that the finite volume corrections are subject to a universal scaling relation.
We also provide a more detailed discussion of the technical details and our 
numerical method.

\section{Theoretical Framework}


\subsection{Lagrangian and Boson-Diboson Amplitude}

\begin{figure}
\centering
 \includegraphics[width=.75\linewidth]{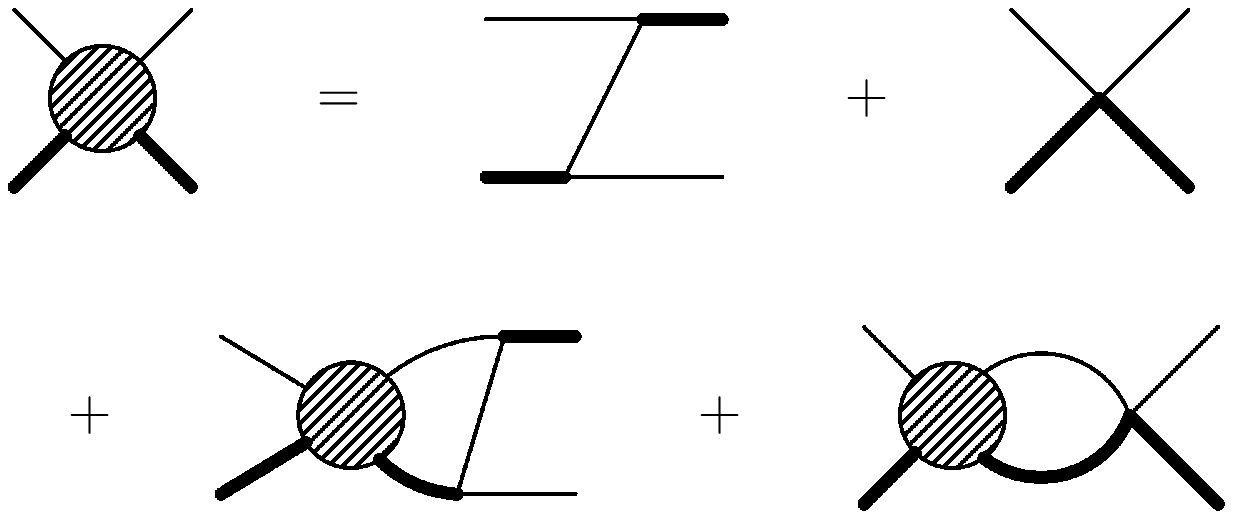}
\caption{Integral/sum equation for the boson-diboson amplitude. The 
narrow lines denote the boson 
propagator and the thick lines denote the full diboson propagator.}
\label{fig_fw_inteq}
\end{figure}

Three identical bosons interacting via short-range forces can be 
described  by the Lagrangian~\cite{Bedaque:1998kg,Braaten:2004rn}
\begin{equation}\label{eq_fw_lagrangian}
\mathcal{L} =
\psi^\dagger \left(i \partial_t + \halb \nabla^2 \right) \psi
+ \frac{g_2}{4} d^\dagger d 
- \frac{g_2}{4}\left(d^\dagger\psi^2 + \mathrm{h.c.}\right)
- \frac{g_3}{36}d^\dagger d \psi^\dagger\psi +\ldots\,,
\end{equation}
where the dots indicate higher order terms. The degrees of freedom in the 
Lagrangian are a boson field~$\psi$ and a non-dynamical
auxiliary field~$d$ representing the composite of two bosons. 
Units have been chosen such that $\hbar=m=1$.
This Lagrangian corresponds to the zero range limit with $r_0=0$. Our
results will therefore be applicable to physical states whose size is large
compared to $r_0$. Corrections from finite range can be included via higher 
order terms in the Lagrangian but will not be considered here.
The coupling constants~$g_2$ and~$g_3$ will be determined by
matching calculated observables
to a two-body datum and a three-body datum, respectively.
The central quantity in the three-body sector of this EFT is 
the boson-diboson scattering amplitude. It determines all three-body 
observables and is given as the solution of an inhomogeneous 
integral equation, depicted diagrammatically in Fig.~\ref{fig_fw_inteq}. A 
detailed discussion of the infinite volume case can be 
found in \cite{Braaten:2004rn}.  Here we follow the strategy of 
Refs.~\cite{Kreuzer:2008bi,Beane:2003da} and focus on the finite volume case.

The system of three bosons is assumed to be contained in 
a cubic box with side length~$L$ and 
periodic boundary conditions. This leads to quantized 
momenta~$\vec{k}=\frac{2\pi}{L}\vec{n},\,\vec{n}\in\mathbb{Z}^3$. 
In particular, the possible loop momenta are quantized. 
As a consequence, the integration over loop momenta is replaced
by an infinite sum.
The divergent loop sums are regulated by a momentum cutoff~$\Lambda$
similar to the infinite volume case.

The finite volume modifies the infrared physics of the system but does 
not change the ultraviolet behavior of the amplitudes. Therefore, the 
renormalization is the same in the infinite and finite
volume cases. Of course, this statement only holds if the momentum 
cutoff~$\Lambda$ is large compared to the momentum scale set by the 
size of the volume, namely~$2\pi/L$, such that the infrared and the 
ultraviolet regime are well separated.
In the numerical calculations presented in this work, the consistent
renormalization of the results will always be verified explicitly.
The two-body sector can be completely renormalized by matching 
the two-body coupling constant~$g_2$ to a low-energy two-body observable, 
namely the two-body scattering length~$a$ or the dimer binding energy. 
Because of the discrete scaling symmetry, the three-body coupling 
approaches an ultraviolet limit cycle. For convenience, the three-body 
coupling constant~$g_3$ is expressed in the 
form~$g_3=-9g_2^2 H(\Lambda)/\Lambda^2$. 
The cutoff dependence of the dimensionless function~$H(\Lambda)$ is then
given by~\cite{Braaten:2004rn}
\begin{equation}\label{eq_fw_H}
H(\Lambda)=\frac{\cos\left[s_0\log(\Lambda/\Lambda_\ast)+\arctan s_0\right]} 
{\cos\left[s_0\log(\Lambda/\Lambda_\ast)-\arctan s_0\right]},
\end{equation}
where $s_0 \approx 1.00624\dots$ and $\Lambda_\ast \approx
2.62 \kappa_*$ is a three-body 
parameter that can be fixed from a trimer binding energy or another 
three-body datum.

\begin{figure}
 \includegraphics[width=0.75\linewidth]{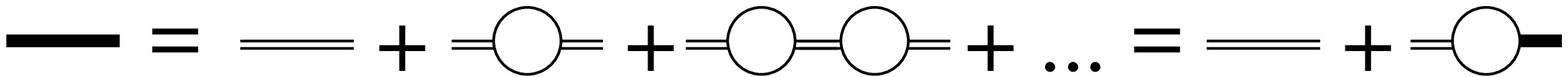}
\caption{The full diboson propagator (thick solid line) is obtained by
dressing the bare diboson propagator (double lines)
with bosonic loops (narrow solid lines).}
\label{fig_fw_dimprop}
\end{figure}

For the diboson lines in the three-body equation depicted in
Fig.~\ref{fig_fw_inteq}, the full, interacting diboson 
propagator~$D$ has to be used. This quantity 
corresponds to the exact two-body scattering amplitude.
It is obtained by dressing the bare diboson propagator, 
given by the constant~$4i/g_2$, 
with bosonic loops as shown in Fig.~\ref{fig_fw_dimprop}. 
This leads to an infinite sum that can be evaluated 
analytically. For a diboson with energy~$E$, the propagator is
\begin{equation}\label{eq_fw_dimprop}
D(E) = \frac{32\pi}{g_2^2} \Bigg[ \frac{1}{a} - \sqrt{-E} + \frac{1}{L}
\sum_{\substack{\vec{\jmath} \in \mathbb{Z}^3\\ \vec{\jmath}\neq 0}}
\frac{1}{|\vec{\jmath}|} \e^{-|\vec{\jmath}|L\sqrt{-E}} \Bigg]^{-1}\,.
\end{equation}
In the limit $L\to\infty$, this expression 
reduces to the full diboson propagator in the infinite volume case.

Having obtained the full diboson propagator, the integral equation for the 
boson-diboson scattering  amplitude can be written down explicitly.
Using the diagrammatical representation in Fig.~\ref{fig_fw_inteq}
and the Feynman rules derived from the effective Lagrangian 
(\ref{eq_fw_lagrangian}), we obtain:
\begin{eqnarray}\label{eq_fw_inteq_0}
\mathcal{A}(\vec{p}, \vec{k}; E) &=& -\frac{g_2^2}{4}
\frac{1}{E-p^2/2-k^2/2-(\vec{p}+\vec{k})^2/2} -\frac{g_3}{36}\\
\nonumber &&\quad + i \int\intdpo{q_0} 
L^{-3}\sum_{\vec{q}\in\frac{2\pi}{L}\mathbb{Z}^3}
\left[\frac{g_2^2}{4}\;\frac{1}{q_0-q^2/2}\;\frac{D(E-q_0+q^2/4)}
{E-p^2/2-q_0-(\vec{p}+\vec{q})^2/2}\right.\\
\nonumber &&\hspace{4.5cm}\left. + \frac{g_3}{36} \frac{1}{q_0-q^2/2}
D(E-q_0+q^2/4)\right]\mathcal{A}(\vec{q}, \vec{k}; E)
\end{eqnarray}
Here, $\vec{p}$ ($\vec{k}$) are the momenta of the incoming (outgoing) 
dibosons, while the momenta of the incoming (outgoing) bosons
are $-\vec{p}$ ($-\vec{k}$). The boson legs have been put on shell
but the diboson legs remain off-shell.
The total energy of the system, $E$, can be treated as a parameter
of the equation. The integration over 
the loop energy can be performed by virtue of the residue theorem. 
This yields the sum equation
\begin{equation}\label{eq_fw_inteq_inhom}
\mathcal{A}(\vec{p}, \vec{k}; E) = -\frac{g_2^2}{4}\mathcal{Z}_E(\vec{p}, 
\vec{k}) +\frac{8\pi}{L^3}\sum_{\vec{q}\in\frac{2\pi}{L}\mathbb{Z}^3}
\mathcal{Z}_E(\vec{p}, \vec{q})
\tau_E(q)\mathcal{A}(\vec{q}, \vec{k}; E),
\end{equation}
with
\begin{align}\label{eq_fw_derf_z}
\mathcal{Z}_E(\vec{p}, \vec{k}) &= \left[\left(p^2+\vec{p}\cdot
\vec{k}+k^2-E\right)^{-1}+\frac{H(\Lambda)}{\Lambda^2}\right],\\
\label{eq_fw_def_tau}
\tau_E(q) &= \bigg[\frac{1}{a}-\sqrt{\frac{3q^2}{4}-E}+\sum_{\substack{
\vec{\jmath} \in \mathbb{Z}^3\\ \vec{\jmath}\neq 0}} \frac{1}{L|\vec{\jmath}|} 
\e^{-|\vec{\jmath}|L\sqrt{\frac{3q^2}{4}-E}} \bigg]^{-1}.
\end{align}

If the energy~$E$ is near a trimer energy~$E_3^{(n)}<0$, the 
amplitude~$\mathcal{A}$ exhibits a simple pole and the dependence on the 
momenta seperates:
\beq\label{eq_fw_pole}
\mathcal{A}(\vec{p}, \vec{k}; E) \longrightarrow \frac{\mathcal{F}
(\vec{p})\mathcal{F}(\vec{k})}{E-E_3^{(n)}}, \quad\mbox{as }E
\rightarrow E_3^{(n)}.
\eeq
Matching the residues on both sides of Eq.~(\ref{eq_fw_inteq_inhom}), 
the bound-state equation
\beq\label{eq_fw_boundstate}
 \mathcal{F}(\vec{p})=
\frac{8\pi}{L^3}\sum_{\vec{q}\in\frac{2\pi}{L}\mathbb{Z}^3}
\mathcal{Z}_E(\vec{p}, \vec{q})\tau_E(q)
\mathcal{F}(\vec{q})
\eeq
is obtained. Values of the energy~$E$, for which this homogeneous sum 
equation has a solution, are identified with the energies~$E_3^{(n)}$ 
of the trimer states.

\subsection{Cubic symmetry}

In the infinite volume case, only s-wave bound states are formed. However, 
in a finite cubic volume, the spherical symmetry of the infinite volume is 
broken to a cubic symmetry. In the language of group theory, the infinitely 
many irreducible representations of the spherical symmetry group~$O(3)$ are 
mapped onto the five irreducible representations of the cubic group~$O$. The 
representations of the spherical symmetry group are now reducible and can 
therefore be decomposed in terms of the five irreducible representations of 
the cubic group. On the other hand, a quantity~$\psi_s$ transforming 
according to the irreducible representation~$s$ of~$O$ can be written in 
terms of the basis functions of the spherical symmetry, i.e. the spherical harmonics~$Y_{lm}$, via
\beq\label{eq_fw_kubic_harmonic}
\psi_s(\vec{r}) = \sum_{\ell, t} R_{\ell t}(r) K_{s\ell t}(\hat{r}) =
\sum_{\ell, t} R_{\ell t}(r) \left[\sum_{m} C_{s\ell m}^{(t)} Y_{\ell m}(\hat{r})\right],
\eeq
where $\hat{r} = \vec{r} / |\vec{r}|$ and $t$~is an additional index needed 
if the representation labeled by~$\ell$ appears in the irreducible 
representation~$s$ more than once. The linear combinations $K_{s\ell t}$ of spherical harmonics are called \lq\lq kubic harmonics''~\cite{BetheVdLage:47}. The values of 
the coefficients~$C_{s\ell m}^{(t)}$ are known for values of~$\ell$ as large 
as~12~\cite{Altmann:65}.

In order to make contact with the infinite volume formalism, 
Eq.~(\ref{eq_fw_boundstate}) is rewritten using Poisson's resummation 
formula. This identity states that an infinite sum over integer vectors may 
be replaced by an infinite sum over integer vectors of the Fourier transform 
of the addend, i.e.
\beq\label{eq_fw_poisson}
\sum_{\vec{n}\in\mathbb{Z}^3}f(\vec{n}) = \sum_{\vec{m}\in\mathbb{Z}^3}
\hat{f}(\vec{m})
\eeq
where $\hat{f}(\vec{m})=\int_{\mathbb{R}^3}\intdd{y}\e^{\ii 2\pi \vec{m}
\cdot\vec{y}}f(\vec{y})$ is the Fourier transform of~$f$. Applying the 
identity~(\ref{eq_fw_poisson}) to Eq.~(\ref{eq_fw_boundstate}) yields
\beq\label{eq_fw_boundstate_p}
 \mathcal{F}(\vec{p}) = \frac{1}{\pi^2}\sum_{\vec{n}\in\mathbb{Z}^3}\int\intdd{y}
                        \e^{\ii L \vec{n}\cdot\vec{y}}\mathcal{Z}_E(\vec{p}, 
                        \vec{y})
                        \tau_E(y)\mathcal{F}(\vec{y}).
\eeq
The term with $\vec{n}=\vec{0}$ gives the corresponding equation in the 
infinite volume, while the other terms of the infinite sum may be viewed 
as corrections due to momentum quantization and the breakdown of the 
spherical symmetry. The explicit recovery of the infinite volume term is 
useful for bound states, since in this case the analytic structure of the 
amplitude, i.e.~the pole at the binding energy, is identical and only the 
position of the pole is changed. This approach may be inappropriate for 
states belonging to the continuous scattering region of the infinite volume.

In order to evaluate the angular integration in Eq.~(\ref{eq_fw_boundstate_p}),
all quantities with angular dependence are expanded in terms of the basis 
functions of the irreducible representations of~$O(3)$, namely in spherical 
harmonics. The amplitude~$\mathcal{F}$ itself is assumed to transform under 
the trivial representation $A_1$ of the cubic group, since the $\ell = 0$ 
representation is solely contained in $A_1$. The amplitude can therefore be 
written as
\beq\label{eq_fw_exp_amp}
\mathcal{F}(\vec{p}) = \sum_{\ell, t} F_{\ell t}(p) K_{A_1 \ell t}(\hat{p}) =
                       \sum^{(A_1)}_{\ell, t} F_{\ell t}(p) 
                          \sum_{m=-\ell}^{\ell} C_{A_1 \ell m}^{(t)} Y_{\ell m}(\hat{p}).
\eeq
The sum runs over those values of~$\ell$ associated with the 
$A_1$-representation of~$O$. The first few values are $\ell=0,4,6,8,\dots$. 
Since the first occasion where an $\ell$-value appears more than once is 
$\ell=12$, the summation over the multiplicity and hence the index~$t$ will 
be suppressed in the following.

The only angular dependence of the quantity~$\mathcal{Z}_E(\vec{p},\vec{y})$ 
is on the cosine of the angle~$\theta_{\vec{p}\vec{y}}$ between~$\vec{p}$ 
and~$\vec{y}$.
Therefore,~$\mathcal{Z}_E(\vec{p},\vec{y})$ can be expanded in Legendre 
polynomials~$P_\ell(\cos\theta_{\vec{p}\vec{y}})$. These polynomials can in 
turn be expressed in spherical harmonics via the addition theorem, yielding
\beq\label{eq_fw_exp_z}
\mathcal{Z}_E(\vec{p},\vec{y}) = \sum_{\ell=0}^\infty Z^{(\ell)}_E(p,y) 
P_\ell(\cos\theta_{\vec{p}\vec{y}})
                               = \sum_{\ell=0}^\infty Z^{(\ell)}_E(p,y) 
\frac{4\pi}{2\ell+1}\sum_{m=-\ell}^{\ell} 
Y_{\ell m}^\ast(\hat{p})Y_{\ell m}(\hat{y}).
\eeq
The exponential function in Eq.~(\ref{eq_fw_boundstate_p}) can be rewritten 
using the identity
\beq\label{eq_fw_exp_exponential}
\e^{\ii L \vec{n}\cdot\vec{y}} = 4\pi\sum_{\ell=0}^\infty \ii^\ell 
j_\ell(L|\vec{n}|y) \sum_{m=-\ell}^{\ell} 
Y_{\ell m}^\ast(\hat{n})Y_{\ell m}(\hat{y}),
\eeq
where $j_\ell$ is the spherical Bessel function of order~$\ell$.

With these expansions, the angular integration in 
Eq.~(\ref{eq_fw_boundstate_p}) can be performed. The integral over the 
three spherical harmonics depending on~$\hat{y}$ yields Wigner 3-$j$~symbols. 
Projecting on the $\ell$th partial wave results in an infinite set of 
coupled integral equations for the quantities~$F_\ell$:
\begin{equation}\begin{split}\label{eq_fw_partialwaves}
F_{\ell}(p) &= \frac{4}{\pi} \int_0^\Lambda \mathrm{d} y\, y^2 \bigg[
Z^{(\ell)}_E(p, y) \tau_E(y) \frac{1}{2\ell+1}F_\ell(y)\\
& \qquad\qquad + 2\sqrt{\pi}\sum_{\substack{\vec{n}\in\mathbb{Z}^3\\ \vec{n}\neq 0}}
\sum_{\lp, \mp}^{(A_1)}\sum_{\lpp, \mpp}
\begin{pmatrix}\lp & \lpp & \ell \\ 0 & 0 & 0\end{pmatrix}
\begin{pmatrix}\lp & \lpp & \ell \\ \mp & \mpp & 0\end{pmatrix}
\frac{C_{\lp\mp}}{C_{\ell 0}}Y_{\lpp \mpp}(\hat{n})\\
& \qquad\qquad\qquad \;\times \sqrt{\frac{(2\lp+1)(2\lpp+1)}{2\ell+1}}\;
\ii^\lpp j_\lpp(L|\vec{n}|y) Z^{(\ell)}_E(p,y) \tau_E(y) F_{\lp}(y) \bigg].
\end{split}\end{equation}
The $\lp$ sum runs over the partial waves associated with the $A_1$ 
representation. The quantity $Z^{(\ell)}_E(p,y)$ can be calculated from 
Eq.~(\ref{eq_fw_exp_z}) to be
\beq\label{eq_fw_zl}
Z^{(\ell)}_E(p, y) = (2\ell+1)\left[\frac{1}{py}\,
Q_\ell\left(\frac{p^2+y^2-E}{py}\right) +
\frac{H(\Lambda)}{\Lambda^2}\delta_{\ell 0}\right].
\eeq
Here, $Q_\ell$ is a Legendre function of the second kind. The three-body 
contact interaction contributes only to the s-wave, as expected.

Since the bound states in the infinite volume are s-wave states, 
Eq.~(\ref{eq_fw_partialwaves}) is specialized to the case~$\ell=0$, yielding
\begin{equation}\begin{split}\label{eq_fw_l=0}
F_0(p)=&\frac{4}{\pi}\int_0^\Lambda \dd y\, y^2 \bigg[Z_E^{(0)}(p, y) 
\tau_E(y) \bigg(1+\sum_{\substack{\vec{n}\in\mathbb{Z}^3\\ \vec{n}\neq 0}}
\frac{\sin(L|\vec{n}|y)}{L|\vec{n}|y}\bigg) F_0(y) \\
&+ 2\sqrt{\pi}\sum_{\substack{\vec{n}\in\mathbb{Z}^3\\ \vec{n}\neq 0}}
\sum_{\lp=4,6,\dots}^{(A_1)}\sum_{\mp=-\lp}^{\lp}\ii^\lp 
j_\lp(L|\vec{n}|y)Y_{\lp\mp}^\ast(\hat{n})Z_E^{(0)}(p,y)\tau_E(y)
C_{\lp\mp}F_\lp(y)\bigg]
\end{split}\end{equation}
The specialization of Eq.~(\ref{eq_fw_zl}) to the case $\ell=0$ reads
\beq\label{eq_fw_z0}
Z_E^{(0)}(p,y) = \frac{1}{2py}\log\left(\frac{p^2+py+y^2-E}{p^2-py+y^2-E}
\right)+\frac{H(\Lambda)}{\Lambda^2}.
\eeq
\begin{table}
\newlength{\tabcolsepold}
\setlength{\tabcolsepold}{\tabcolsep}
\setlength{\tabcolsep}{10pt}
\begin{tabular}{|*{4}{c|}}
\hline
~ & \multicolumn{3}{c|}{$\sum_{\substack{\vec{n}\in\mathbb{Z}^3\\ |\vec{n}| = n}} \halb\left(Y_{\ell m}+Y_{\ell, -m}\right)$}\\
\cline{2-4}
$n$ & $\ell = 0, m=0$ & $\ell = 4, m=0$ & $\ell = 4, m=4$ \\
\hline\hline
101  & 606   & 96.23    & 81.33    \\
1001 & 9126  & $-$94.05 & 20.69    \\
3501 & 39678 & 118.67   & $-$15.73 \\
\hline
\end{tabular}
\setlength{\tabcolsep}{\tabcolsepold}
\caption{Integer vector sums over real combinations of $Y_{\ell m}$ for a given 
absolute value $n$ (\lq\lq angular sums''). For higher partial waves, the value 
of the sum is smaller and the contribution of the partial wave is suppressed.}\label{tab:ylmsum}
\end{table}
The second line of Eq.~(\ref{eq_fw_l=0}) indicates admixtures from higher 
partial waves. Since the leading term in the expansion of the spherical 
Bessel function is $1/(L|\vec{n}|y)$, these contributions are suppressed by 
at least~$a/L$. They will therefore be small for volumes not too small 
compared to the size of the bound state. Moreover, contributions from higher 
partial waves will be suppressed kinematically for shallow states with small 
binding momentum. This is ensured by the spherical harmonic in the second 
line of Eq.~(\ref{eq_fw_l=0}). For higher partial waves, the angular sum yields smaller 
prefactors relative to the s-wave ($\ell=0$)
for a given absolute value~$n$ (see Table~\ref{tab:ylmsum}). 
Only for small lattices, i.e. when $a/L$ is 
large, this behavior is counteracted by terms stemming from the spherical 
Bessel function and higher partial waves may contribute significantly. The 
first numerical calculations presented in this work were therefore mainly performed 
neglecting contributions from higher partial waves. For two specific examples, however, 
we explicitly calculate the contribution from the next partial wave $\ell=4$
and demonstrate that it is small. Some details on the
numerical solution of Eq.~(\ref{eq_fw_l=0}) are given in the Appendix.
In the following section, we present our results.

\section{Results and discussion}
By employing the formalism laid out in the previous section, we have 
calculated energy levels in finite cubic volumes of varying side lengths. 
The results of these calculations are presented in the following.
For convenience, the dependence of the energies on the boson mass 
$m$ is reinstated in this
section.

\subsection{Positive scattering length}
We first present results for systems with $a>0$. In this regime, a physical 
diboson state with a binding energy~$E_D=-1/(ma^2)$ exists. This energy is 
therefore identical with the threshold for the break-up of a trimer into a 
diboson and a single boson. We choose states with different energies in the 
infinite volume, including shallow as well as deeply bound states:
\begin{itemize}
\item[Ia:] $E_3^\infty\, = -1.18907/(ma^2)$, \quad $\Lambda_\ast a=5.66$,
\item[Ib:] $E_3^\infty\, = -27.4427/(ma^2)$, \quad $\Lambda_\ast a=5.66$,
\item[Ic:] $E_3^\infty\, = -9440.91/(ma^2)$, \quad $\Lambda_\ast a=5.66$,
\item[II:] $E_3^\infty\, = -5.04626/(ma^2)$, \quad $\Lambda_\ast a=1.66$,
\item[III:] $E_3^\infty\,= -11.1322/(ma^2)$, \quad $\Lambda_\ast a=3.66$,
\end{itemize}
Here, $E_3^\infty$ is the trimer energy in the infinite volume. Note that 
the states Ia, Ib and Ic appear in the same physical system chracterized 
by $\Lambda_\ast a=5.66$. For each of these states, its energy in the finite 
cubic volume has been calculated for various values of the box side 
length~$L$. In order to check the consistency of our results, the calculation 
was carried 
out for several cutoff momenta~$\Lambda$. For each cutoff, the three-body 
interaction parameterized by~$H(\Lambda)$ has been adjusted such that the 
infinite volume binding energies are identical for all considered values 
of~$\Lambda$. If our results are properly renormalized, the results for the 
different cutoffs should agree with each other up to an uncertainty of order 
$1/(\Lambda a)$ stemming from the finiteness of the cutoff.

\begin{figure}[t]
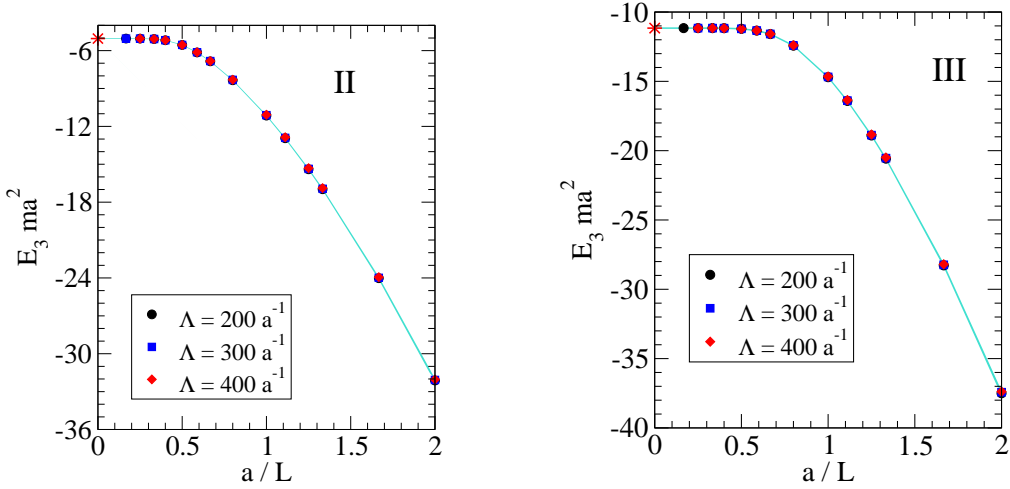

  \centerline{
    \includegraphics*[width=5.83cm]{fig_new.eps}\qquad\qquad
    \includegraphics*[width=6cm]{fig_nunu.eps}
  }
  \caption{Variation of the trimer energy $E_3$ with
           the side length $L$ of the cubic volume for the states II
           (left) and III (right). Plotted are three datasets for
           different values of the cutoff parameter $\Lambda$, together
           with the $1/(\Lambda a)$ bands. The point $a/L=0$ corresponds to 
           the infinite volume limit.}
  \label{fig:s23}
\end{figure}

The results for the states~II and~III are depicted in Fig.~\ref{fig:s23} 
for box sizes between $L=6a$ and $L=a/2$. The values obtained for different 
cutoffs indeed agree with each other within the depicted uncertainty bands. 
Note that these bands do not represent corrections from higher orders of the 
EFT.
For both states, the infinite volume limit is smoothly approached. As the
volume becomes smaller, the energy of the states is more and more
diminished. This corresponds to an increased binding with decreasing box size.

In the infinite volume, state~III is more deeply bound than state~II. 
Na\"{\i}vely, one therefore expects the former to have a smaller spatial 
extent than the latter. The size of the state can be estimated via the 
formula~$(-mE_3^\infty)^{-1/2}$, yielding $0.45a$ for state~II and $0.3a$ 
for state~III. Hence, a given finite volume should affect state~II more 
strongly than the smaller state~III. This behavior can indeed be observed. 
For example, considering a cubic volume with side length $L=2a$, the energy 
of state~II deviates 10\% from the infinite volume value, whereas the 
corresponding difference for state~III is less than one percent. On the 
other hand, the box size for which the energy shift of the 
state~III amounts to 10\% is roughly $1.3a$.

For both states, we can now form the dimensionless number 
$r = -mE_3^\infty L_{10\%}^2$, where~$L_{10\%}$ is the box size at which 
the energy differs by 10\% from the infinite volume value. This yields 
$r \approx 20$ for state~II and $r \approx 19$ for state~III. The 
approximate equality of the two values of $r$ may 
indicate the presence of universal scaling in the finite volume version 
of the Effective Theory.

\begin{figure}[t]
  \centerline{
    \includegraphics*[width=6cm]{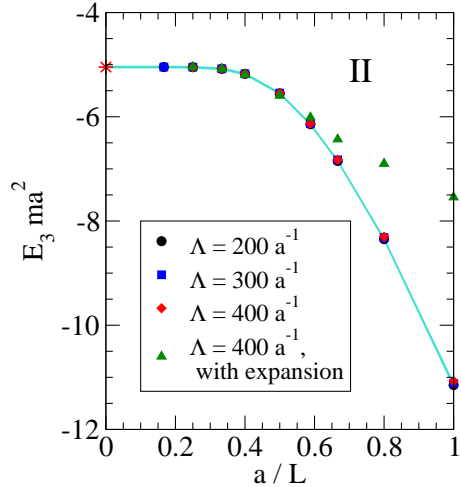}
  }
  \caption{Variation of the trimer energy $E_3$ with
           the side length $L$ of the cubic volume for state II.
           Plotted are three datasets for different values of the cutoff 
           parameter $\Lambda$, together
           with the $1/(\Lambda a)$ bands, and one dataset obtained using 
           a Taylor expanded version of
           the integral kernel.}
  \label{fig:s2exp}
\end{figure}

In Fig.~\ref{fig:s2exp}, the three datasets obtained for state~II are 
plotted again, this time in comparison to the results from a calculation 
using the expansion of the kernel used in Ref.~\cite{Kreuzer:2008bi}
and described in Appendix~\ref{sec:expansion}. For 
large values of the box size~$L$, the results of both calculations agree 
with each other. For volumes smaller than~$1.7a$, the result of the 
calculation with the expansion deviates from the result of the full 
calculation. For this size of the volume, the full result differs by 
about~20\% from the infinite volume binding energy. Accordingly, the 
expansion employed for the integral kernel can not be applicable any 
longer. In Fig.~\ref{fig:s2exp}, results are only depicted for volumes 
with $L \ge a$. For smaller volumes, the results of calculations using 
the expansion become cutoff dependent, and are hence not are not 
properly renormalized anymore.

\begin{figure}[t]
  \centerline{
    \includegraphics*[width=8cm]{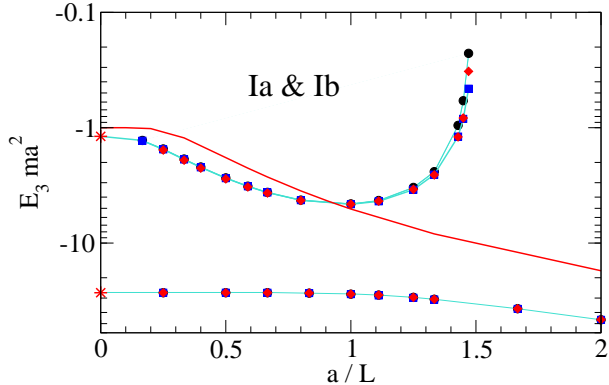}
  }
  \caption{Variation of the trimer energy $E_3$ with
           the side length $L$ of the cubic volume for the
           states Ia (upper curve) and Ib (lower curve).
           Plotted are three datasets for different values of the
           cutoff parameter $\Lambda$, together with the $1/(\Lambda a)$ bands 
           (circles:~$\Lambda = 200\,a^{-1}$;
            squares:~$\Lambda = 300\,a^{-1}$;
            diamonds:~$\Lambda = 400\,a^{-1}$).
           The solid line depicts the diboson energy.}
  \label{fig:s1ab}
\end{figure}

In Fig.~\ref{fig:s1ab}, we show our results for the two states~Ia and~Ib in the same physical system characterized by $\Lambda_\ast a = 5.66$. The volume dependence of the two states is again shown for box sizes between $L=6a$ and $L=a/2$. The curve corresponding to the more deeply bound state~Ib shows a behavior similar to the one observed for the states~II and~III. The binding energy remains constant until the volume is small enough to affect the state. At this point, the energy of the state is more and more diminished as the volume becomes smaller.

The behavior of the shallow state~Ia is different. In the region $L \approx a$, the binding is not further increased. The results for smaller volumes show a sharp rise of the three-body energy. The energy of the state becomes positive near $L=0.67a$. State~Ia is close to the threshold for boson-diboson scattering 
in the infinite volume located at $E_D=-1/(ma^2)$. For comparison, 
we calculated the energy of the physical diboson according to~\cite{Beane:2003da}. The resulting curve is the solid line in Fig.~\ref{fig:s1ab}. Like the energy of the three-body bound states, the energy of the diboson is diminished in finite volumes. For volumes of the size $L\approx 1.2a$, the energy of the 
three-body state~Ia becomes larger than the diboson energy and starts to grow.
This behavior is consistent with the observation that states are always shifted away from the threshold in a finite volume. In the two-body sector, for example, continuum states have been shown to have a power law dependence on the volume, while the volume dependence of bound states is dominated by exponentials~\cite{Beane:2003da}. The data shown for state~Ia can be explained by an exponential for $L>a$, which characterizes the state as a bound state. For $L<a$, the data is consistent with a power law, indicating the state indeed behaves like a boson-diboson scattering state if its energy is above the diboson energy.

The other investigated states do not show such a transition since their energy is well below the diboson energy for all considered volumes. It is unclear whether other states would show a similar behavior for smaller box sizes. If this is not the case, the transition from bound to unbound would occur only for states with infinite volume binding energies up to a critical value. For a conclusive analysis of the nature of the described transition, more data is still needed. It would be interesting to see whether such a transition shows up in lattice data for a state that is very close to the diboson threshold in the infinite volume.

\begin{table}[tb]
\begin{tabular}{|c|c|c|c|c|}
\hline
$\quad L/a\quad$ & $\quad\Lambda a\quad$ & $\quad E_3(L)\,ma^2$, 
$\quad\ell=0,4\quad$ & $\quad E_3(L)\,ma^2$, $\quad\ell=0\quad$ & 
$\quad E_2\,ma^2\quad$\\
\hline\hline
1.25 & 200 & $-$4.30392 & $-$4.24545 & $-$3.53099\\
\hline
     & 200 & $-$4.57097 & $-$4.57753 & \\
\cline{2-4}
1    & 300 & $-$4.60298 & $-$4.60872 & $-$5.07581\\
\cline{2-4}
     & 400 & $-$4.59579 & $-$4.60056 & \\
\hline
0.9  & 200 & $-$4.31223 & $-$4.27927 & $-$6.068 \\
\hline
\end{tabular}
\caption{Energies $E_3(L)$ of state~Ia for box sizes near $L=a$ calculated 
with and without admixture of the $\ell=4$ amplitude. The dimer energy~$E_2$ 
is shown for comparison.}\label{tab:hpw}
\end{table}
Since the size of the finite volumes where the state crosses the diboson energy is comparable to the size of the state itself, the breaking of the spherical symmetry may be a relevant effect here. 
To assess the influence of the higher partial waves, we extract from 
Eq.~(\ref{eq_fw_partialwaves}) 
two coupled equations for the s-wave amplitude~$F_0$ 
($\ell=0$) and the $\ell = 4$~amplitude $F_4$. 
These coupled equations are then solved in a coupled channel approach.
The energies of state~Ia obtained by this method are summarized and compared to the s-wave only results in Table~\ref{tab:hpw}. For~$L=1.25a$, the state is still below the dimer state. The inclusion of the higher partial wave leads to
a small downward shift in the energy.
For~$L=a$, we have done calculations using 3 different cutoffs. 
The results for different cutoffs agree to two significant digits
indicating the results are properly renormalized, but the binding is 
slightly reduced by the $\ell=4$ contribution.
For $L=0.9a$, the effect of the higher partial wave is again a small downward 
shift. All results show only a deviation of about 1\% from the s-wave 
only result. 
In summary, we find that the correction from the $\ell=4$ admixture 
is extremely small. Moreover, the correction is of the same order of magnitude as 
the finite cutoff corrections in our calculation and a more quantitative
statement requires improved numerical methods.
\begin{table}[tb]
\begin{tabular}{|c|c|c|c|}
\hline
$\quad L/a\quad$ & $\quad\Lambda a\quad$ & $\quad E_3(L)\,ma^2$, 
$\quad\ell=0,4\quad$ & $\quad E_3(L)\,ma^2$, $\quad\ell=0\quad$\\
\hline\hline
1   & 200 & $-$11.86 & $-$11.15 \\
\cline{2-4}
    & 400 & $-$11.79 & $-$11.08 \\
\hline
0.7 & 200 & $-$19.06 & $-$20.70 \\
\cline{2-4}
    & 400 & $-$18.97 & $-$20.64 \\
\hline
\end{tabular}
\caption{Energies $E_3(L)$ of state~II for box sizes $L=a$ and $L=0.7a$ calculated with and
	  without admixture of the $\ell=4$ amplitude.}\label{tab:s2hpw}
\end{table}
In order to establish an estimate of typical corrections from higher partial waves, we have also performed 
calculations including the $\ell=4$ contributions for the more deeply-bound
state~II. The resulting numbers are given in Tab.~\ref{tab:s2hpw}. The investigated box sizes
are about three times larger than the state itself. The contribution of the higher partial wave is 
now several percent. This is still a small correction but considerably larger than the finite cutoff uncertainty. This 
suggests that the extremely small corrections for state Ia
are related to its unusual behavior. The dominance of the s-wave my be associated with the closeness 
of the state to the threshold.
A more detailed investigation of higher partial waves including a description of the numerical methods will 
be the subject of a future publication.

Since the shallower state~Ia is more affected by a given finite volume than the deeper state~Ib, the ratio of the energies of the states is changing. In the infinite volume, this ratio is 23.08. For $L=1.5a$, just before the shallow state crosses the dimer energy, this ratio has decreased to 7.4. Note that this ratio differs from the discrete scaling factor $\exp(2\pi/s_0)\approx 515$ even in the infinite volume limit. This behavior is expected for shallow states close to the bound state threshold~\cite{Braaten:2004rn}. The ratio~$\exp(2\pi/s_0)$ will be approached when deeper states are considered. For example, the infinite volume ratio of state~Ib and the much more deeply bound state~Ic is~344 and already closer to the discrete scaling factor 515.


If we assume that the combination $r = -mE_3^\infty L_{10\%}^2$ is indeed a universal number, we are able to predict~$L_{10\%}$ for the states~Ia and~Ib. The results are $L_{10\%} \approx 0.85a$ for state~Ib and $L_{10\%} \approx 4a$ for state~Ia. The energies calculated for these volumes are $E_3(L=0.85a) \approx 29.1 / (ma^2)$, corresponding to an 8\% shift, for state~Ib and $E_3(L=4a) \approx 1.285 / (ma^2)$, also corresponding to an 8\% shift. These findings support the assumption that the finite volume corrections obey universal scaling relations.

The states~Ia and~Ib appear in the same physical system characterized by $\Lambda_\ast a = 5.66$. In this system, an even more deeply bound state, denoted as state~Ic, is present. In the infinite volume, the energy of this state is $E_3^\infty = -9401.32/(ma^2)$. This corresponds to a binding momentum of~$97a^{-1}$. Since this is already comparable to the momentum cutoffs of a few hundred
inverse scattering lengths 
employed before, we also performed calculations for this state using a much larger cutoff of~$4500~a^{-1}$. The three-body force for this cutoff has been fixed such that the energy of the most shallow state~Ia is reproduced. The resulting energy of state~Ic is then $-9440.91/(ma^2)$. This differs from the energy obtained using the smaller cutoff by 0.4\%. This difference can be attributed to effects stemming from the finiteness of the cutoff. We have calculated the energy of this state in finite volumes of the sizes $L=a$, $L=0.75a$ and $L=0.5a$. The results of this calculations are summarized in Table~\ref{tab:s1c}. The values obtained using the large cutoff~$\Lambda a=4500$ show no effect of the finite volume at all. From the infinite volume binding energy, the size of the state can be estimated via~$(mE_3^\infty)^{-1/2}$ to be~$0.01a$. So, we do not expect any visible effect since the finite volume is fifty times as large as the state itself. However, for the smaller cutoff~$\Lambda a = 400$, there are very small deviations from the infinite volume energy. But these deviations are smaller than the uncertainty stemming from the finiteness of the cutoff, which is estimated to be of order~$1/(\Lambda a) = 0.25\%$.

\begin{table}[tb]
\begin{tabular}{|c|cc|cc|}
\hline
  & \multicolumn{2}{c|}{$\Lambda a = 4500$} &\multicolumn{2}{c|}{$\Lambda a = 400$}\\\hline
$\quad L/a\quad$ & $\quad E_3(L)\,ma^2\quad$ & $\quad\delta_{rel}\quad$ 
& $\quad E_3(L)\,ma^2\quad$ & $\quad\delta_{rel}\quad$ \\
\hline\hline
$\infty$ & $-$9440.91 & --  & $-$9401.32 & -- \\
1        & $-$9440.91 & 0\% & $-$9401.36 & $\approx 10^{-6}$ \\
0.75     & $-$9440.91 & 0\% & $-$9400.53 & $\approx 10^{-5}$ \\
0.5      & $-$9440.91 & 0\% & $-$9399.38 & 0.02\% \\
\hline
\end{tabular}
\caption{Energy~$E_3$ of state~Ic for different volume sizes~$L$ and two different cutoffs~$\Lambda$. The relative deviation from the infinite
volume value
$\delta_{rel}=(E_3-E_3^\infty)/E_3^\infty$ is also given.}\label{tab:s1c}
\end{table}

\subsection{Negative scattering length}
Now we turn to systems with $a<0$. In this regime, the two-body interaction is attractive but no diboson bound state exists in the infinite volume limit. 
The only possible breakup process for a three-boson bound state is therefore the breakup into three single bosons. The threshold for this process is $E=0$. As for the case with positive scattering length, we choose states with different energies in the infinite volume:
\begin{itemize}
\item[NI:] $E_3^\infty\,= -9/(ma^2)$, \quad\phantom{5} $\Lambda_\ast a=5.55$,
\item[NII:] $E_3^\infty\, = -4/(ma^2)$, \quad\phantom{5} $\Lambda_\ast a=4.23$,
\item[NIII:] $E_3^\infty\, = -0.2/(ma^2)$, \quad $\Lambda_\ast a=2.60$,
\end{itemize}
Here, $E_3^\infty$ is the trimer energy in the infinite volume. For each of these states, its energy in the finite cubic volume has been calculated for various values of the box side length~$L$. In order to check the consistency of our results, the calculation was carried out for several cutoff momenta~$\Lambda$. For each cutoff, the three-body interaction parameterized by~$H(\Lambda)$ has been adjusted such that the infinite volume binding energies are identical for all considered values of~$\Lambda$. 

\begin{figure}[t]
  \centerline{
    \includegraphics*[width=5.83cm]{fig_nine.eps}\qquad\qquad
    \includegraphics*[width=6cm]{fig_four.eps}
  }
  \caption{Variation of the trimer energy $E_3$ with
           the side length $L$ of the cubic volume for the states NI
           (left) and NII (right). Plotted are three datasets for
           different values of the cutoff parameter $\Lambda$, together
           with the $1/(\Lambda a)$ bands. The point $a/L=0$ corresponds to 
           the infinite volume limit.}
  \label{fig:sn1n2}
\end{figure}

\begin{figure}[t]
  \centerline{
    \includegraphics*[width=6cm]{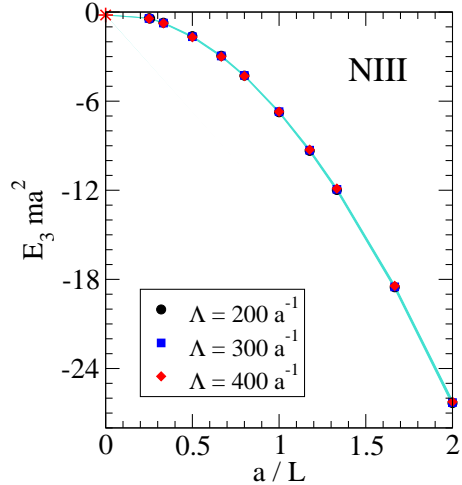}
  }
  \caption{Variation of the trimer energy $E_3$ with
           the side length $L$ of the cubic volume for the states NIII.
           Plotted are three datasets for different values of the cutoff parameter $\Lambda$,
           together with the $1/(\Lambda a)$ bands.}
  \label{fig:sn3}
\end{figure}

The results for the states~NI and~NII are shown in Fig.~\ref{fig:sn1n2} for box sizes between $L=2a$ and $L=a/2$. The results for the state~NIII with box sizes between $L=8a$ and $L=a/2$ are depicted in Fig.~\ref{fig:sn3}.
The values obtained with different cutoffs all agree with each other 
within the uncertainty bands indicating proper renormalization. 
Our findings are similar to those in the positive scattering length regime. All three states smoothly approach the infinite volume limit. As the box size becomes smaller the energy of the state is more and more diminished. The overall behavior is identical to the one in the positive scattering length case described in the previous section.

\begin{table}
\begin{tabular}{|l|*{5}{c|}}
\hline
\makebox[1.3cm][l]{\,State} & \makebox[3cm][c]{$E_3^\infty\,ma^2$} & \makebox[3cm][c]{$E_3(L=a)\,ma^2$} & \makebox[2.5cm][c]{$\delta_{rel}$} & \makebox[2.5cm][c]{$L_{10\%} / a$} & \makebox[2.5cm][c]{$L_{100\%} / a$} \\
\hline\hline
\,NI   & $-9$   & $-11.6$ & 29\%   & 1.2 & 0.7 \\
\,NII  & $-4$   & $-8.7$  & 118\%  & 1.8 & 1.05 \\
\,NIII & $-0.2$ & $-6.7$  & 3250\% & 7.2 & 4.2 \\
\hline
\end{tabular}
\caption{For the states NI, NII and NIII, the infinite volume energy~$E_3^\infty$ and the energy in a finite volume with side length $L=a$ are shown together with the relative deviation from the infinite volume value $\delta_{rel}=(E_3-E_3^\infty)/E_3^\infty$. Also given are the box sizes leading to an energy shift of 10\% and 100\%, respectively.\label{tab:neguniv}}
\end{table}

The more deeply bound a state is in the infinite volume, the 
smaller is its spatial extent. Estimating the size via the 
formula~$(-mE_3^\infty)^{-1/2}$ yields $a/3$ for state~NI, $a/2$ 
for state~NII, and $2.2a$ for state~NIII. A given finite volume should 
therefore affect state~NII more than state~NI, and state~NIII should be the 
most affected. When considering a cubic volume with side 
length $L=a$, we find the energies given in Table~\ref{tab:neguniv}. The relative deviation of state~NI is indeed four times smaller than the shift for state~NII and a hundred times smaller than the shift for the shallow state~NIII. On the other hand, the box length~$L_{10\%}$, for which the energy of each states deviates~10\% from its infinite volume value, is smaller the more deeply bound a state is. These values are also given in Table~\ref{tab:neguniv}.

As in the case of positive scattering length, we have 
investigated the dimensionless combination~$r=-mE_3^\infty L_{10\%}^2$. 
From state~NI, we obtain $r \approx 13$. This leads to the predictions $L_{10\%} = 1.8a$ for state~NII and $L_{10\%} = 8a$ for state~NIII
in good agreement with the explicitly calculated values.
Additionally, we form another dimensionless combination, 
$r^\prime = -mE_3^\infty L_{100\%}^2$, where $L_{100\%}$ is the 
box side length where the energy of the state is twice as large 
as the infinite volume value. For the three investigated states, 
this box size is also given in Table~\ref{tab:neguniv}.
From the value for the state~NI, we get $r^\prime \approx 4.4$. 
From this value, we predict $L_{100\%} = 1.05a$ 
for state~NII and $L_{100\%} = 4.7a$ for state~NIII. 
Again, the value for state~NII is as predicted, 
whereas the value for state~NIII is about 10\% off the prediction from
universal scaling.

For states far away from the threshold, we expect that the regimes of negative and positive scattering length are governed by the same scaling factor. Therefore, the dimensionless combination $r$ should, for such deeply bound states, have a common value for both signs of~$a$. For positive scattering lengths, state~Ib is an example of a rather deeply bound state. For this state, we find $L_{10\%} = 0.79a$ and from that $r=16.8$. For negative scattering lengths, we chose a state with $E_3^\infty = -27/(ma^2)$ by setting $\Lambda_\ast a = 8.54$. The energy of this state is shifted by 10\% in a volume with side length~$0.73a$, yielding $r=14.4$. The values of $r$ for the two signs of the scattering length are indeed close to each other for these two states. This behavior provides numerical evidence for the universality of finite volume effects for both positive and negative scattering lengths. 

\section{Conclusions}

In this paper, we have extended our earlier studies of the Efimov spectrum
in a cubic box with periodic boundary conditions \cite{Kreuzer:2008bi}.
The knowledge of the finite volume modifications of the spectrum
is important in order to
understand results from future 3-body lattice calculations. 
Using the framework of EFT, we have derived a general set
of coupled sum equations for the Efimov spectrum in a finite volume. 

Specializing to $\ell = 0$, we have 
calculated the spectrum for both positive and negative scattering lengths
and verified the renormalization in the finite box explicitly.
We have removed the expansion for small finite volume shifts
used in Ref.~\cite{Kreuzer:2008bi} and presented an extension that 
can be applied to arbitrarily large shifts. Typically, the binding of all
three-body states increases as the box size is reduced. Moreover,
we provided a more detailed discussion of the technical details and our 
numerical method.
We have investigated the breakdown of the linear approximation in detail 
and find that the expansion is applicable as long as the 
finite volume shift in the energy is not larger than 15--20\%
of the infinite volume energy.

We have studied the spectrum for positive and negative scattering lengths in
detail and provided numerical evidence for universal scaling of the finite
volume effects. The scaling properties can be quantified by the 
dimensionless product $r$ of the three-body 
binding energy in the infinite volume limit and the square of the box 
length corresponding to a finite volume shift of ten percent of the 
infinite volume energy. For sufficiently deep states, we obtained 
numerical evidence that $r$ approaches a universal number for 
both signs of the scattering length. These findings suggest that 
the properties of deeper states in the finite volume can be obtained 
from a simple rescaling and do not require explicit calculations. A more 
detailed analysis of this issue along the lines of~\cite{Braaten:2002sr} is in progress.

For positive scattering lengths, we have investigated a spectrum of three
states in the same physical system and studied the behavior of the 
shallowest state in the vicinity of the dimer energy which specifies
the scattering threshold in the infinite volume. We found that this 
state drastically changes its behavior as a function of the 
box length once its energy becomes equal to the dimer energy.
At this point the energy of the shallowest three-body state starts to grow
and eventually becomes positive.  The observed behavior is consistent with
exponential suppression of the finite volume corrections below the dimer
energy and power law suppression above.

The effect of an admixture of the $\ell=4$ partial wave has been investigated
for two different three-body states
for box sizes of the order of the size of the state. 
For a generic state well separated from threshold, we found corrections of the order of a few percent for 
volumes that are about three times larger than the state itself. For the state closest to threshold, the 
effect of this admixture turns out to be surprisingly small for volumes 
about twice as large as the state and was found to be 
less than $1\%$.
As the finite cutoff corrections for this state are about the same size,
a more quantitative study requires an improved treatment of these corrections.
Such studies are in progress.

Finally, our method should be extended to the three-nucleon system.
This requires also the inclusion of higher
order corrections in the EFT and finite temperature effects
as lattice calculations will inevitably be performed at a small,
but non-zero, temperature. Work in these directions is in progress.
With high statistics lattice QCD simulations of three-baryon systems 
within reach \cite{Beane:2009gs}, the calculation of the 
structure and reactions of light nuclei appears now feasible in the 
intermediate future. Our results demonstrate that the finite
volume corrections for such simulations are calculable and
under control. This also opens the possibility to
test the conjecture of an infrared limit cycle in QCD for quark masses
slightly larger than the physical values \cite{Braaten:2003eu}.

\begin{acknowledgments}

We would like to thank D. Lee and B. Metsch for helpful discussions.
This research was supported by the DFG through
SFB/TR 16 \lq\lq Subnuclear structure of matter'' and the BMBF
under contracts No. 06BN411 and 06BN9006.

\end{acknowledgments}

\appendix

\section{Numerical treatment}

In this appendix, we give some details on the numerical solution of
Eqs.~(\ref{eq_fw_partialwaves}, \ref{eq_fw_l=0}).
The starting point for a first numerical treatment of the formalism is 
the homogeneous integral equation
\beq\label{eq_fw_num_start}
F_0(p)=\frac{4}{\pi}\int_0^\Lambda \dd y\, y^2 Z_E^{(0)}(p, y) \tau_E(y) \bigg(1+\sum_{\substack{\vec{n}\in\mathbb{Z}^3\\ \vec{n}\neq 0}}
\frac{\sin(L|\vec{n}|y)}{L|\vec{n}|y}\bigg) F_0(y).
\eeq
The first step is to transform this equation into a finite-dimensional 
problem. Due to the oscillatory nature of the integrand, it is not sensible 
to use a finite number of sampling points. Therefore, a set of basis functions 
will be used. The choice of the basis functions is guided by the knowledge of 
the bound-state amplitude in the infinite volume case. Asymptotically, the 
amplitude in the infinite volume behaves 
like~\cite{Bedaque:1998kg,Braaten:2004rn}
$$F_0(p) \rightarrow \frac{1}{p}\cos\left(s_0\log(p/p_\ast)\right),$$
with the universal number~$s_0=1.00624\dots$ and a momentum scale~$p_\ast$. 
The bound-state amplitude obtained by the infinite volume formalism has 
precisely such a form, with only a few zeros in the 
interval~$\left[0;\Lambda\right]$ 
(See, e.g., Ref.~\cite{Bedaque:1998kg}). Therefore, the basis functions~$\xi_i$
are chosen as Legendre functions~$P_i$ with logarithmic arguments as
\beq\label{eq_fw_basis}
\xi_i(p) = \frac{1}{p}P_i(2\log_2(p/\Lambda + 1) - 1).
\eeq
These basis functions are orthogonal with respect to a suitably chosen 
scalar product:
\beq\label{eq_fw_scp}
\int_0^\Lambda \dd p\, \frac{p^2}{p+\Lambda} \xi_i(p) \xi_j(p) = 
(\log 2)\,\frac{1}{2i+1}\delta_{ij}.
\eeq
The amplitude~$F_0$ in~Eq.~(\ref{eq_fw_num_start}) is replaced by its 
expansion in the basis functions $\sum_{j=0}^\infty f_j \xi_j(p)$ and the 
$i$th component is projected out using the scalar product. The resulting 
equation
\beq\label{eq_fw_matrix_inf}
\frac{\log 2}{2i+1} f_i = \sum_{j=0}^\infty K_{ij}(E) f_j,
\eeq
with
\beq\label{eq_fw_matrix_def}
K_{ij}{(E)}=\frac{4}{\pi}\int_0^\Lambda \dd y\, y^2 
\bigg[\int_0^\Lambda \dd p\, \frac{p^2}{p+\Lambda} \xi_i(p) Z_E^{(0)}(p, y) 
\bigg]\tau_E(y) \bigg(1+\sum_{\substack{\vec{n}\in\mathbb{Z}^3\\ 
\vec{n}\neq 0}} \frac{\sin(L|\vec{n}|y)}{L|\vec{n}|y}\bigg) \xi_j(y),
\eeq
can be interpreted as a matrix equation when truncating the set of basis 
functions.

A non-trivial solution of Eq.~(\ref{eq_fw_matrix_inf}) can be found if the 
condition
\beq\label{eq_fw_determinant}
\det\left[K_{ij}{(E)} - \frac{\log 2}{2i+1}\delta_{ij}\right] = 0
\eeq
is fulfilled. Values of~$E$, for which this is the case, are then considered 
as trimer energies in the finite volume. This is only a valid interpretation 
if the truncation of the basis and the neglect of the higher partial waves 
induce only small corrections for the result.

\label{sec:expansion}
The value of the parameter~$E$ for which the 
condition~(\ref{eq_fw_determinant}) is satisfied is found via a root 
finding algorithm. This requires the calculation of~$K_{ij}(E)$ in each 
iteration. To save numerical effort, it is possible to expand the kernel 
around the binding energy in infinite volume if the  shift in the binding 
energy is small. The expansion is done up to first order:
\beq\label{eq_fw_expand}
K_{ij}(E) \approx K_{ij}(E^\infty) + \left.
\frac{\dd K_{ij}}{\dd E}\right|_{E=E^\infty}\cdot(E - E^\infty),
\quad\mbox{ for } \frac{E-E^\infty}{E^\infty} \ll 1.
\eeq
This expansion has been used to obtain the results in~\cite{Kreuzer:2008bi}. 
It is applicable as long as the 
finite volume shift in the energy is not larger than 15--20\%
of the infinite volume energy.

In the following, details on the numerical methods used to calculate the 
kernel matrix elements as defined in Eq.~(\ref{eq_fw_matrix_def}) are 
presented. The integration over~$p$ is performed using a logarithmically 
distributed Gauss-Legendre quadrature with 64~points.

The integrand of the $y$-integration is strongly oscillating due to the 
$\sin(L|\vec{n}|y)$ term. Therefore, this integral is evaluated using the 
Fast Fourier Transformation technique (FFT). The computation of Fourier-type 
integrals via FFT is explained in detail in~\cite{NumRep:07}. The sampling 
for the FFT is the most time-consuming part of the calculation. The number 
of points to sample, $M$, is determined by the highest ``frequency'' that is 
to be accessed. The frequencies in the present case are the values 
of~$L|\vec{n}|$. The highest possible frequency is connected to $M$ and 
the integration range~$\Lambda$ by
\beq\label{eq_fw_freqhigh}
L|\vec{n}_\mathrm{max}|\,\frac{\Lambda}{M} < \pi 
\quad\Rightarrow\quad
M > \frac{L|\vec{n}_\mathrm{max}|\Lambda}{\pi}.
\eeq
For realistic values of these parameters, namely box sizes of a few~$a$, 
an~$|\vec{n}_\mathrm{max}|$ of about 3000 and cutoff values of several 
hundred~$a^{-1}$, a typical value for~$M$ is $2^{20}$ or about one million 
sampled points.

\begin{figure}[t]
  \centerline{
    \includegraphics*[width=8cm]{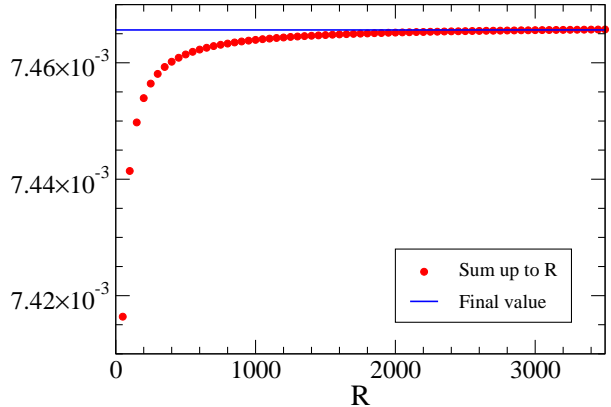}
  }
  \caption{Convergence of the sum over vectors in $\mathbb{Z}^3$
   as a function of the radius $R$.}
  \label{fig:sum}
\end{figure}

A summation over three-dimensional integer vectors of a quantity depending 
only on the absolute values involves 
substantial double counting when na\"{\i}vely done. 
To circumvent this problem, 
an ordered list with all absolute values that 
are possible for vectors in~$\mathbb{Z}^3$ and their multiplicity has been 
created. The result of the summation can be viewed as converged when going 
to absolute values of about $R\approx 5000$ (see the example
in Fig.~\ref{fig:sum}). 
For radii $R$ from about~2000 on, the 
intermediate sums oscillate around the converged result. To reduce runtime, 
the result of the summation has been calculated as a mean of 15 intermediate 
sums for radii from~2750 to~3500.

The calculation of the matrix elements has been performed on the cluster of 
the HISKP at the University of Bonn. 
As already stated above, the number of sampling points for the FFT is the 
parameter with the largest influence on the runtime. When using the expanded 
integral kernel, the matrix and the derivative matrix have to be calculated 
only once. For a typical sample size of $2^{20}$, this takes about 
100~minutes. When going to smaller volumes, the shifts in the binding energy 
become larger and the Taylor expansion of the integral kernel breaks down. 
In this case, it is inevitable to recalculate the kernel matrix in each 
iteration of the root finding algorithm. This amounts to~10 to~15 evaluations 
of the kernel matrix and a runtime of~8 to~10~hours per data point.

\end{document}